\documentstyle[]{l-aa}
\input psfig.sty
\begin{document}
\thesaurus{     03     
              (11.09.1 NGC 7172  
               11.19.1       
               13.25.2)}     

\def\NH{N_{\rm H}}
\def\NHp{N_{\rm H p}}
\def\e{\rm e}
\title{The variable X/$\gamma$-ray spectrum of the Seyfert 2
 galaxy NGC~7172} 

\author{ F.~Ryde\inst{1} \and J.~Poutanen \inst{2}
\and  R.~Svensson \inst{1}\and S.~Larsson \inst{1}\and S.~Ueno \inst{3}
} 

\institute{Stockholm Observatory, S-133 36 Saltsj\"obaden, Sweden
\and  Uppsala Astronomical Observatory, Box 515, S-751 20, Uppsala, Sweden
\and  Kyoto University, Sakyo-ku Kyoto, Japan}

\offprints{F.~Ryde}

\date{Received ; accepted}

\maketitle

\begin{abstract}
      
A broad band X-, gamma-ray spectral study of the Seyfert 2 galaxy NGC 7172 
is presented. We use our {\it ASCA} observations from May 1995 
and combine these with the {\it CGRO} OSSE data  from March
1995. The only Seyfert 2 galaxy
 previously to have been modelled over such a broad spectral range 
is NGC 4945.

The {\it ASCA} GIS data alone, can be acceptably described by a
model consisting of a  power law, with a photon index $ \Gamma = 
1.58 \pm 0.12$, affected by absorption due to intervening neutral 
matter along the line of sight, with  an equivalent hydrogen 
column density $\NH = (8.2^{+0.6} _{-0.5}) \times 10^{22}$cm$^{-2}$.
 An Fe K$\alpha$ emission line is not required
by the fits.  The observed  flux in the $2-10$ keV range is $F_{2-10} = 
(4.75\pm 0.09) \times 10^{-11}$ erg cm$^{-2}$ s$^{-1}$, which corresponds
to a small increase since the {\it{Ginga}} measurement in October
1989. The power law found is flatter than what is expected
in the unified model of Seyfert galaxies.

Combining these {\it ASCA} data with the {\it CGRO} OSSE data 
from March 1995, the above conclusions do not change. In this case, we also
 find 
that the most probable model for the data is an absorbed
power law, but being affected by a high energy exponential cut-off. 
The power law is still flat with $\Gamma = 
1.54 \pm 0.13$, while $\NH = (8.1 \pm 0.6 ) 
\times 10^{22}$ cm$^{-2}$.
The OSSE data do not contradict
the flat spectrum found from using the {\it ASCA} data alone,
 which could have been an artifact of the
narrow spectral range covered by this single instrument.  

Even though it is not possible to completely reject an alternative
explanation to the observed flat spectrum in terms of a transmission 
or a reflection model, the spectral index of the underlying power
law of NGC 7172
appears actually to have varied from $1.85$ to $1.5$ 
since the {\it Ginga}
observations in 1989.  The e-folding energy is relatively well 
constrained and lies at  $140 ^{+310} _{-70}$ keV, using the {\it CGRO}
OSSE viewing period with the highest quality data. We note, however, 
that the {\it CGRO} OSSE spectral shape appears to be variable on a
time scale of weeks.


\end{abstract}

\section{Introduction}

In the unified model of Seyfert galaxies  (Antonucci \cite{antonucci}), the
observed differences between Seyfert 1 and Seyfert 2 galaxies are
explained as an orientation effect. All Seyfert galaxies are suggested
to be intrinsically the same type of object, having a central,
accreting, massive black hole, around which there is a luminous disk 
and hot gas. A molecular torus surrounds the central X-ray source,
obscuring it for edge-on viewing, but not for face-on
viewing within
the opening angle of the torus. Seyfert galaxies viewed from an angle
at which the central region is directly visible are classified as 
Seyfert 1,
and when the central region is obscured, they are classified as
Seyfert 2. The model is based on the observations of polarized 
broad lines in the 
optical spectra of several Seyfert 2 galaxies (see, e.g., Antonucci
\& Miller  \cite{AntMill}) and the presence of strong absorption 
in the X-ray spectra of Seyfert 2 galaxies, indicating a
molecular torus (e.g., Koyama et al. \cite{Koyama}). 

Some Seyfert 2 galaxies have been observed to have an intrinsic spectrum
which is flatter than has been observed for Seyfert 1 galaxies
(Smith \& Done \cite{SmithDone}, Cappi et al. \cite{Cappi}). This contradicts
the unified model, which predicts the same average intrinsic photon index 
for Seyfert 1 and Seyfert 2 galaxies. 
However,
observations with a  single satellite are usually limited to a fairly 
narrow spectral range, e.g., $0.5 - 10$ keV for {\it ASCA}. 
If the radiation
from the object undergoes extensive absorption, this range may be
too narrow to make it possible to determine whether the 
intrinsic spectrum is  really  flat or whether it is intrinsically steep 
and maybe affected by a complex absorber and/or Compton reflection. 
By combining data from two or more 
satellites, one can broaden the spectral coverage and reduce these 
uncertainties. In this paper we conduct such an analysis on the
Seyfert 2 galaxy NGC 7172. The only Seyfert 2 galaxy previously 
studied over a broader band using non-simultaneous 
{\it ASCA}/{\it Ginga}/{\it CGRO} OSSE data
is NGC 4945 (Done et~al. \cite{MadDone}). 
The average, non-simultaneous {\it Ginga}/{\it CGRO} OSSE spectrum 
of three Seyfert 2 galaxies has also been studied (Zdziarski et~al. 
\cite{Z5}).

An X-ray source detected by the sky surveys of
{\it ARIEL} V/SSI and  {\it HEAO}-1/A2, was later
identified as NGC 7172, which is an edge-on, Sa(pec)\footnote{Anupama 
et~al. (\cite{Anupama}) advocate a morphological type S0-Sa for the galaxy.} 
spiral galaxy
(de Vaucouleurs et~al. \cite{RC3} (RC3)). The galaxy
has a strong equatorial dust lane and an equivalent neutral
hydrogen column density of about $4 \times 10^{21}$ atoms cm$^{-2}$,
determined from  the Balmer decrement, as compared to the galactic 
 column density of $1.7 \times 10^{20}$ cm$^{-2}$. NGC 7172
lies at a redshift of $z = 0.0086$, corresponding to a distance 
of 51.8 Mpc ($H_0 = 50 $ km s$^{-1}$ Mpc$^{-1}$), in the
southern compact group HCG 90 (see Sharples et~al. \cite{Sharples}; 
Anupama et~al. \cite{Anupama}). The X-ray spectrum 
was studied by {\it EXOSAT} in October 1985 (Turner \& Pounds 
\cite{TurnerPounds}),
by {\it Ginga} in October 1989 (Warwick et~al. \cite{Warwick};
Nandra \& Pounds \cite{NandraPounds};
Smith \& Done \cite{SmithDone}), 
by {\it ROSAT}/PSPC in November 1992 (Polletta et~al. \cite{Polletta}),
by {\it CGRO} OSSE in March 1995, and
by {\it ASCA} on 13 May, 1995 (Ryde et~al. \cite{Ryde}) and on
 17 May, 1996 (Matt et~al. \cite{Matt}). Here we analyse and fit 
the March/May 1995 data, as well as the 1989 {\it Ginga} data.
The {\it ASCA} data for NGC 7172
are extracted using XSELECT (ftools v3.6) and all the data analysis is
performed with XSPEC v8.5 (Arnaud  \cite{Arnaud}).

\section{The {\it Ginga} spectrum of NGC 7172 }

In order to compare with earlier results, we also analyse the 
{\it{Ginga}} data, starting with a simple model spectrum, consisting of
a power law, absorption due to intervening neutral matter along the line
of sight,  and an iron K$\alpha$ emission line with the line energy fixed
at $6.4$ keV and the line width at $0.1$ keV. In agreement with Warwick et
al. (\cite{Warwick}), we find a photon spectral index $\Gamma = 1.85\pm
0.04$, and an equivalent hydrogen 
column density $\NH = (10.5\pm0.5) \times 10^{22}$ cm$^{-2}$ with $\chi
^2 = 30.3$, for the number of degrees of freedom,
d.o.f. $ = 23$ (reduced $\chi ^2 \equiv \chi ^2 _{\nu} = 1.32 $). The model
(absorption $\times$ power law) $\propto e ^{- \sigma (E) \NH}
 E ^{- \Gamma}$, where $\sigma ( E )$ is the photo-electric
 absorption cross section, 
will be referred to as `model P'. All errors given, unless otherwise 
stated, are
$90$ $\%$ confidence intervals for one interesting parameter, 
i.e., $\Delta \chi^2 = 2.7$. The {\it Ginga} data show a
hardening of the spectrum above 10 keV, which can be interpreted as 
an indication of a Compton reflection component. Such a component can be
expected when, for instance, the relatively cold matter in an 
accretion disk subtends
a substantial solid angle as seen from the X- and $\gamma$-ray 
emitting corona. The amount of reflection is measured by the 
parameter $R$, which is proportional to the ratio of the observed 
reflection spectrum and the observed underlying continuum. $R$
 is normalized in such a way that $R = 1$ for reflection by an 
infinite disk illuminated by an isotropic continuum source. 
Including a Compton
reflection component in the model (XSPEC model pexrav by
Magdziarz \& Zdziarski \cite{magdziarz}) with
the cosine of the inclination angle = 0.45, one finds
$\Gamma = 1.98 ^{+0.10} _{-0.12}$ and  $\NH = (11.0\pm0.5) 
\times 10^{22}$ cm$^{-2}$
with an amount of reflection, $R = 0.8 ^{+0.7} _{-0.5}$, and $\chi ^2 /
$d.o.f. $ = 23.1 / 22 $ ($\chi ^2 _{\nu} =  1.05$). The model
absorption $\times$ (power law + reflection), will be referred to as
`model R'.   The hardening may also be explained by 
a transmission model, in which the absorption is approximated with a 
 dual absorber.
 Here, two column 
densities are used, $\NH$ having
complete coverage, and $\NHp$ covering only a fraction, $C _ F$, of 
the source. The transmission $\propto (1-C _ F) \Phi _ {E}(\NH) 
+  C _ F \Phi _{E} (\NH + \NHp) $, where the function $\Phi _{E}$ is
 derived
in a 1D, two-stream approximation, taking into 
account both photoelectric absorption and Thomson scattering,
allowing the absorber to be marginally Thomson thick.
Such a model can be a crude description of a non-uniform 
density distribution of absorbing matter. Using this `model T' 
on the {\it Ginga}
data one finds  $\Gamma = 2.00 ^{+0.15} _{-0.00}$ and the column densities
$\NH = (10.0 _{-4.4} ^{+0.5}) \times 10^{22}$ cm$^{-2}$ and $\NHp = (70
^{+330} _{-35}) \times 10^{22}$ cm$^{-2}$ with $C _ F = 0.32 ^{+0.17} 
_{-0.23}$. The fit had  $\chi ^2 /
$d.o.f.$ = 25.7 / 22 $ ($\chi ^2 _{\nu} = 1.17$).

The {\it Ginga} flux of NGC 7172 in the 2 - 10 keV range, $ F_{2-10} = 
4.1 \times 10^{-11}$ erg cm$^{-2}$ s$^{-1}$.
 
\section{The  {\it ASCA} spectrum of NGC 7172 }

We observed NGC 7172 with the {\it ASCA} satellite 
from 23:18 UTC on the 12 May
until 16:00 UTC on the 13 May 1995, with its four telescopes, which
focus the X-rays onto the two CCD detectors, SIS0 and SIS1 and onto the
two gas scintillation proportional counters, GIS2 and GIS3.

\subsection{The GIS data}
 
In screening the GIS data, standard, conservative criteria are applied: 
Minimum angle to Earth limb $= 10^\circ$, angle 
from bright Earth (illuminated limb) $ = 20^\circ$, minimum cut-off
rigidity $ = 6$ GeV/c. The spatial region filter, centred around 
the object, is chosen to have a radius of $\sim 6'$. Background 
spectra are accumulated from regions 
of the observation itself, which are sufficiently far away to avoid 
contamination from the source. The final results do not differ
significantly, if, instead, the background is extracted from
the event files for a blank sky background provided by the  {\it ASCA} Guest
Observer Facility. Furthermore, they are not very sensitive to the screening 
criteria adopted. The extracted data are regrouped to contain a minimum 
of $20$ counts per channel to allow the use of the $\chi^2$-test. 
The spectrum clearly shows a soft ($ < 2$ keV) excess above the extrapolation
of the higher energy continuum found below. These
channels, however, do not have many counts after the background subtraction 
and they are ignored in the modelling.
The net exposure time for both detectors is $27$ ks.

\subsubsection{GIS analysis and results}

  The GIS2 and GIS3 observations are presented in Table 1, where model
P, i.e., a power law and a neutral absorber, has been
fitted for a first 
comparison. The observations are in agreement with each other and 
we analyse the data by fitting the GIS2 and the GIS3 spectra
simultaneously. We start off by studying model P, without an Fe K$\alpha$
line. The fit is satisfactory with 
 $\Gamma = 1.58\pm0.12$ and an
equivalent hydrogen column density, $\NH = (8.2^{+0.6} _{-0.5}) \times
 10^{22}$ 
cm$^{-2}$, with $\chi ^2 /$d.o.f. $ = 316 / 360  $ ($\chi ^2 _{\nu} = 
0.88$). Even though the fit does not require the model to have
 any additional components, we try adding an Fe K$\alpha$ line, described by
three additional parameters,
which reduces the $\chi ^2$ by 5 units, $\chi ^2 /$d.o.f. $ = 311 
/ 357 $ ($\chi ^2 _{\nu} = 0.87 $).
By studying the confidence contour plot of the Fe line strength versus
the line energy, it is obvious that the Fe line is not significantly
detected and we do not add it in the following fits. The equivalent width
is $\sim $60 eV with a 90 $\%$ upper limit of 120 eV.

 The observed energy flux in the $2 - 10$ keV range is $F_{2-10} = 4.75
 \times 10^{-11}$ erg cm$^{-2}$ s$^{-1}$, which after absorption correction
  becomes $F_{2-10}^{\rm corr}= 7.5 
\times 10^{-11}$ erg cm$^{-2}$ s$^{-1}$.
The luminosity, with $z = 0.0086$, $H_{0} = 50$ km s$^{-1}$
Mpc$^{-1}$, $ q_{0} = 0$, becomes $L_{2-10}
= 1.52 \times 10^{43}$ erg s$^{-1}$.

As a reflection component is suspected to exist in the {\it Ginga}
data, we continue by trying model R with a free 
Compton reflection parameter.
However, for the GIS data the form of the fitted model does not change
appreciably, with the reflection parameter, $R$, remaining close to zero
and having an upper limit of 3.5.
In the next step of the analysis, we try fitting model T to the data.
 The covering 
fraction cannot be constrained  by the data.
The GIS data do not reach sufficiently high in energy to be 
able to constrain the parameters for these models.
We therefore adopt the simple power law model as the best fit model for the 
{\it ASCA} GIS data.

Figure 1 shows the {\it ASCA} GIS2 data, together with the
folded, best-fit model.

\begin{table*}

\caption[]{The four {\it ASCA} detectors fitted with `model P' (power law 
and a neutral absorber). 
All errors given are
$90$ $\%$ confidence intervals for one interesting parameter, 
i.e., $\Delta \chi^2 = 2.7$}

\begin{flushleft}

\begin{tabular}{lcccc}  

\\
\hline
\\

{Detector} & {$\Gamma$}  & {$\NH$} & {$F_{2-10}$}  & { $\chi ^2 / $d.o.f. 
($\chi ^2 _ \nu $)} \\
   &     & {($10^{22}$cm$^{-2}$)}  & {($10^{-11}$ erg 
cm$^{-2}$ s$^{-1}$)}& \\
   
\\
\hline
\\
SIS0 &  1.43$\pm$0.16 &  8.1$\pm$0.7    & 5.1  & 185  / 152 (1.21) 
\\
SIS1 &  1.11$\pm$0.16 & 7.0$\pm$0.8    & 5.5  & 121  / 134 (0.90) 
\\
GIS2  &  1.51$\pm$0.18  & 7.6$\pm$0.9   & 4.8  &  147 / 166 (0.89)  
\\
GIS3 &   1.63$\pm$0.16 &  8.5$\pm$0.7   & 4.7  & 167 / 191 (0.87) 
\\
\\
\hline 
\end{tabular}
\end{flushleft}
\end{table*}

\begin{figure}
\psfig{figure=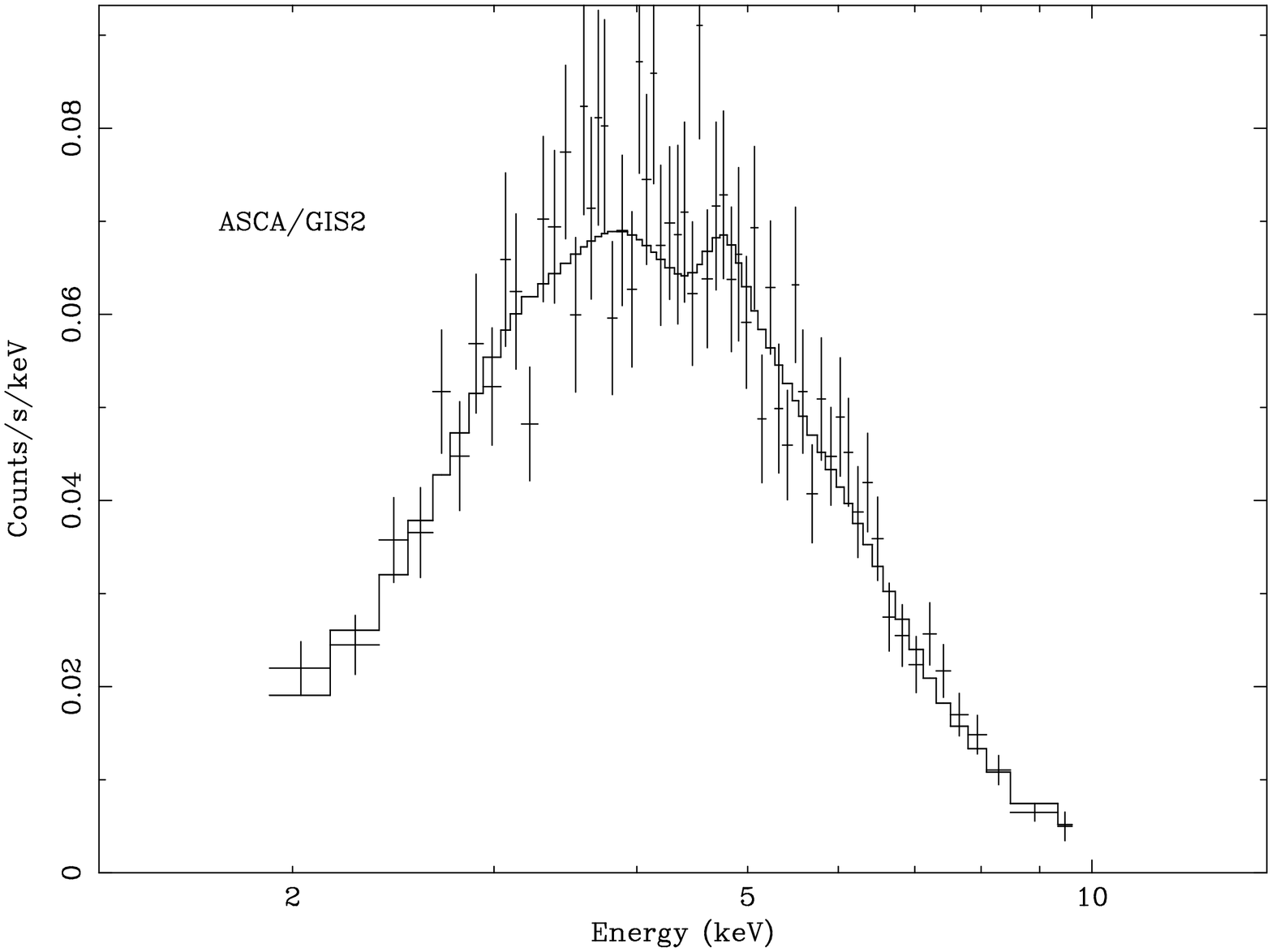,height=6.8truecm,width=8.7truecm,angle=270}
\caption[ ]{ {\it ASCA} GIS2 spectrum of NGC 7172 fitted by `model P' 
     (power law and a neutral absorber)}
\end{figure}

\begin{figure}
\psfig{figure=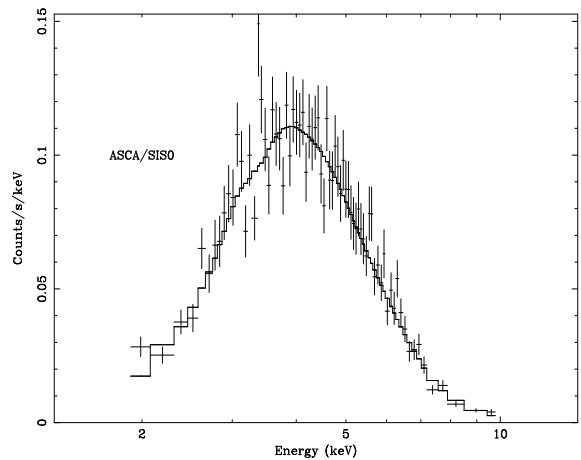,height=6.8truecm,width=8.7truecm,angle=270}
\caption[ ]{{\it ASCA} SIS0 spectrum of NGC 7172 fitted by `model P'
                  (power law and a neutral absorber)}
\end{figure}

\subsection{The SIS data}

The SIS data are sensitive to the screening criteria and give 
different results depending on the criteria adopted.  We therefore 
 apply the standard
conservative data filters: Minimum angle to Earth limb $= 10^\circ$, angle 
from bright Earth (illuminated limb) $ = 20^\circ$, minimum cutoff
rigidity $ = 6$ GeV/c. Only the SIS events that are graded as $0,2,3,4$
are used and the spatial region filter, centred around the object, is 
chosen to have a radius of $\sim 4'$. The background is extracted from
the event files for a blank sky background provided by the ASCA Guest
Observer Facility (November 1994). These files contain events from 3 
deep field observations, without any obvious sources and with a total 
exposure time of about $120$ ks. If, instead, the background is 
extracted from the observation itself, the results do not change
more than a few per cent and the 
conclusions of the subsequent modelling are in principle the same.
 The SIS detectors operated in the 2 CCD
bright mode and had a net exposure time of 26 ks. 
The extracted data
are regrouped  to contain a minimum of $20$ counts per channel 
to allow the use of the $\chi^2$-test. The
low energy channels up to $\sim 1.5$ keV are ignored.

\subsubsection{SIS analysis and results}

The SIS0 data, together with the folded model P fit, are shown in Fig. 2.
The results of this fit and the corresponding ones for SIS1,
are given in Table 1. As shown in the table, the two SIS detectors
give noticeably different results. For instance, the photon indices are
not consistent with each other at the 
confidence level used. From the SIS results themselves, one can not judge 
which observation is the one mainly causing the discrepancy.
We continue by studying the SIS data by fitting the SIS0 and the SIS1 
spectra individually.

Using model R, with a free reflection parameter, the resulting fits give 
a steeper photon index and a large reflection parameter, which is an
indication for a hard tail; for SIS0, $\Gamma = 2.08\pm0.17$ and
$R = 7^{+12} _{-6}$, and for SIS1, $\Gamma = 1.53^{+0.18} _{-0.17} $ 
and $R = 5^{+6}$, only having an upper limit. The
column densities are $\NH = (8.9\pm0.9) \times 10^{22}$ cm$^{-2}$
 and $\NH = (7.6\pm0.9) \times 10^{22}$ cm$^{-2}$, respectively, and the 
fits have $\chi ^2 / $d.o.f.$ = 180 / 151 $ ($\chi ^2 _{\nu} =  1.19$) and
$\chi ^2 / $d.o.f.$ = 118 / 132 $ ($\chi ^2 _{\nu} =  0.897$). 
Here too the iron line is weak, which could pose a problem for the
scenario with reflection off cold, neutral material, where one would
expect a strong iron line. This supports the interpretation of the 
spectral hardening in terms of the transmission model.  

We now apply the transmission model T. The best fit for the SIS0 detector,
gives $\Gamma = 1.9 ^{+0.5} _{-0.4}$ and column  densities of the fully and 
partially covering components, $\NH = (7.7 ^{+1.0} _{-1.2}) \times 10^{22}$ 
cm$^{-2}$  and 
$\NHp = (30^{+45}_{-23}) \times 10^{22}$ cm$^{-2}$, with a covering
fraction, $C _ F = 0.48^{+0.20}_{-0.33}$. The fit has $\chi ^2 /
$d.o.f.$ = 179 / 150 $ ($\chi ^2 _{\nu} =  1.19$). The corresponding fit 
for the SIS1 detector gives  $\Gamma = 1.54 ^{+0.5} _{-0.4}$,  $\NH = 
(6.9^{+0.9} _{-1.0}) \times 10^{22}$ cm$^{-2}$, $\NHp = (75^{+110}_{-15}) 
\times 10^{22}$ cm$^{-2}$, $C _ F = 0.68^{+0.25 }_{-0.45}$ and $\chi ^2 /
$d.o.f.$ = 116 / 132 $ ($\chi ^2 _{\nu} =  0.88$).
The large uncertainty in the partial 
absorber is due to the fact that it is effective only for a narrow 
energy interval at
high energies and therefore only a few data points can constrain the
values.
This model is of the same quality as the reflection model,
and one cannot tell from the data which is the better one.
A more complicated model, using a power law and a dual absorber as well
as Compton reflection, does not increase the quality of the fit. 

\subsection{Conclusions}

The SIS data can constrain models R and T better than the GIS data. 
However, due to the uncertainties in the reduction, where the 
SIS data are sensitive to the screening criteria and give 
different results depending on the criteria adopted, we rely more on the GIS 
results. The SIS0 and the SIS1 
detectors also give results that differ from each other, while
the GIS data are more consistent with each other and are usually
less sensitive to details in the reduction.  In the subsequent investigation,
 we therefore use only
the GIS2 and the GIS3 data for the {\it ASCA} observation.

The errors in the GIS flux can be found by varying the model parameters to
map out the confidence region. The errors then turn out to be of the same 
order as those due to counting statistics.  The estimated energy flux
becomes $F_{2-10} = (4.75 \pm 0.09)
 \times 10^{-11}$ erg cm$^{-2}$ s$^{-1}$, which after absorption correction
becomes $F_{2-10}^{\rm corr}= (7.5 \pm 0.2)
\times 10^{-11}$ erg cm$^{-2}$ s$^{-1}$ and finally the luminosity becomes 
$L_{2-10} = (1.47 \pm 0.03) \times 10^{43}$ erg s$^{-1}$.

 The fitted GIS spectrum is rather flat, $\Gamma = 1.58\pm0.12$, which is in
the lower end of the range found in Seyfert galaxies, having a mean value of 
 $\Gamma = 1.9 - 2.0$  (e.g., 
Nandra \& Pounds \cite{NandraPounds}, Nandra et~al. \cite{Nandraetal}).
It has also varied since the {\it{Ginga}} observations in October 
1989, when $\Gamma = 1.85\pm0.04$, and $ F_{2-10} = 
4.1 \times 10^{-11}$ erg cm$^{-2}$ s$^{-1}$.
The X-ray  spectrum of NGC 7172 has thus become flatter between 1989 and 
1995, and the $2 - 10$ keV flux level has increased somewhat.
The {\it ASCA} observation of NGC 7172 from May 1996 also shows a flat 
spectrum, $\Gamma \sim 1.5$ (Matt et~al. \cite{Matt}).

NGC 7172 did not show any significant variability during the observations,
with the variations being less than 10 $\%$ of the mean count rate.

\section{The {\it CGRO} OSSE spectrum of NGC 7172}

NGC 7172 was observed by {\it CGRO} OSSE during viewing periods (VP) 411.5,
412.0 and 413.0 from 21 February to 21 March 1995. The data
could be affected by the proximity of PKS 2155-304, as both
objects had to be observed 
together, due to the field of view of the OSSE instrument being a few
times larger than the angular separation between the two objects. During
VP 411.5 and 413.0 a two-pointing strategy 
was adopted, where one pointing was centred on NGC 7172 and the second 
was pointed off source, closer to PKS 2155-304. This strategy makes it 
possible to study the contamination of the PKS object, by fitting for the 
two sources, using both viewing positions, with the analysis package IGORE 
(Johnson et~al. \cite{John93}). Such a study was 
conducted by the {\it CGRO} Science Support Center (SSC) and 
resulted in the conclusion that there is no evidence 
for emission from PKS 2155-304 in the OSSE band (Niel Johnson, 1997, 
private communication), so therefore we attribute 
all the flux observed to NGC 7172.

The OSSE data from all pointings were appropriately prepared for 
XSPEC and made available to us by the SSC.
The results from fitting the data with a power law  are shown in 
Table 2.  Viewing periods 412.0 and 413.0 are consistent with each other, 
while the observations from viewing period 411.5 differs significantly,
having a much flatter spectrum. This indicates that NGC 7172 is 
variable in the OSSE range over a time scale 
of a few weeks. VP 413 is the viewing period with the best 
signal-to-noise ratio, 
having the longest effective observation time. Fitting all the 
viewing periods together gives a photon 
index of $2.0 ^{+0.4} _{-0.3}$ and $\chi ^2 /$d.o.f. $ = 142.7 / 150 $
($\chi ^2 _{\nu} = 0.95$).   The fits should be compared 
to those of other Seyfert galaxies observed in the OSSE range
(Johnson et~al. \cite{Johnson}), for which the mean index is $2.4\pm0.25$.

\begin{table}

\caption[]{The {\it CGRO} OSSE observations of NGC 7172, sorted by
viewing periods. The data are fitted with a power law. 
All errors given are
$90$ $\%$ confidence intervals for one interesting parameter, 
i.e., $\Delta \chi^2 = 2.7$}

\begin{flushleft}

\begin{tabular}{lccc}  

\\
\hline
\\

{Viewing} & Detector & {$\Gamma$}  & { $\chi ^2 / $d.o.f. 
($\chi ^2 _ \nu $)} 
\\
{Period} & seconds &   &   
\\
\\
\hline
\\
411.5 & $1.37 \times 10^5$ &  $1.20^{+0.41} _{-0.34}$ & 60.3  / 51 (1.18) 
\\
412.0   & $1.42 \times 10^5$ &  $3.11^{+1.34} _{-1.13}$ & 27.3  / 44 (0.62) 
\\
413.0   & $2.20 \times 10^5$ &  $2.43^{+0.50} _{-0.47}$  & 34.1  / 51  (0.67)  
\\
All   & $4.99 \times 10^5$ &  $1.97 ^{+0.38} _{-0.33}$ & 142.7 / 150 (0.95)
\\
\\
\hline 
\end{tabular}
\end{flushleft}
\end{table}

\section{The combined X/gamma-ray spectrum of NGC 7172}

We now analyse the combined spectrum using the {\it Ginga},
{\it ASCA} GIS, and {\it CGRO} OSSE data discussed above.
As the spectrum of NGC 7172, observed with {\it ASCA} in 1995, is flatter 
than that from {\it Ginga} in 1989, 
it is of interest to combine
the GIS data with the OSSE data to form a broader 
spectral range. The aim is to be able to determine 
the photon index unambiguously 
and thus prove that the slope really has changed and is  not only an
artifact of the narrowness of the spectral range studied. The 
observations were not
simultaneous, but were made only two months apart.

\subsection{The combined {\it ASCA}/OSSE spectrum} 

We start by studying the combined {\it ASCA} GIS and {\it CGRO}
OSSE data. As shown above, the OSSE data alone indicated some 
variability during the couple of weeks it was gathered. We therefore study
different combinations of the GIS and the sets of OSSE data. We use 
the model P, with a free e-folding energy, $E_{\rm fold}$; $E ^{- \Gamma} 
e^{-E / E_{\rm fold}}$, and a single neutral absorber.
We let the GIS and the OSSE 
data have free normalizations compared to each other. 
The results are presented in Table 3.

\begin{table*}
\caption[]{Parameters for the best fits of the models dicussed in the text. 
Unless otherwise stated, all errors given are
$90$ $\%$ confidence intervals for one interesting parameter, 
i.e., $\Delta \chi^2 = 2.7$. Numbers within brackets are 90 $\%$ lower limits.
$^a$OSSE data from VP 411.5 are used.
$^b$OSSE data from VP 412.0 are used.
$^c$OSSE data from VP 413.0 are used. 
$^d$OSSE data from VP 411.5, 
VP 412.0 and VP 413.0 are used.
$^e$Errors are $68$ $\%$ confidence
intervals for one interesting parameter, i.e., $\Delta \chi^2 = 1.0$}
\begin{flushleft}
\begin{tabular}{lcccccccc}  

\\
\hline

\\

{$\#$} & {Model} & {$\Gamma$}  & {$E_{\rm fold}$} &{$\NH$} & {$\NHp$} 
& {$ C _ F $} & {$R$} & { $\chi ^2 / $d.o.f.  ($\chi ^2 _ \nu $)} \\
 &  &       &  (keV)  &{($10^{22}$cm$^{-2}$)}  &{($10^{22}$cm$^{-2}$)}
 & & & \\

\\
\hline
\\
1 & {\it Ginga} P &  1.85$\pm$0.04 & - & 10.5$\pm$0.5 & - & - & - 
 & 30.3 /23 (1.32) \\

2 & {\it Ginga} R &  1.98$^{+0.10} _{-0.12}$ & -  & 11.0$\pm$0.5 & - & - & 
0.8$^{+0.7} _{-0.5}$ & 23.1/22 (1.05) \\

3 & {\it Ginga} T &  2.00$^{+0.15} _{-0.00}$ & -  & 10.0$^{+0.5} _{-4.4}$ & 
70$^{+330}_{-35}$ & 0.32$^{+0.17} _{-0.23}$ & 
 -  & 25.6/22 (1.17) \\

4 & {\it ASCA} P &  1.58$\pm$0.12 & - & 8.2$^{+0.6} _{-0.5}$ & - & -  & - & 
316/360 (0.88) \\

5 & {\it Ginga}$+$OSSE P  $^a$  &  1.84$\pm$0.04 & (780)  &
10.5$\pm$0.4  & - & - 
&   -      & 96.6/74 (1.31) \\

6 & {\it Ginga}$+$OSSE P  $^b$   &  1.84$^{+0.04} _{-0.10}$ & (95) &
 10.5$^{+0.5 } _{-0.4}$  & - & - 
&    -     & 60.6/66 (0.92) \\

7 & {\it Ginga}$+$OSSE P  $^c$   &  1.82$^{+0.05} _{-0.06}$ & (120) &
 10.5$\pm$0.5  & - & - 
&     -    & 66.1/74 (0.90) \\

8 & {\it Ginga}$+$OSSE P  $^d$   &  1.84$^{+0.04} _{-0.05}$ & (350) &
 10.5$\pm$0.4  & - & - 
&      -   & 173.2/173 (1.00) \\

9 & {\it ASCA}$+$OSSE P $^a$& 1.55$\pm 0.11$  &  (550) &
8.0$^{+0.6} _{-0.4}$ & - & - & -  & 378.9/411 (0.92) \\

10 & {\it ASCA}$+$OSSE P $^b$& 1.44$^{+0.19} _{-0.26}$  & 45$^{+510}
_{-25}$ &
8.0$\pm 0.6$ & - & - & -  & 344.9/404 (0.85) \\

11 & {\it ASCA}$+$OSSE P $^c$& 1.54$\pm$ 0.13  & 140$^{+310} _{-70}$ &
8.1$\pm0.6$ & - & - & -  & 351.1/411 (0.85) \\

12 & {\it ASCA}$+$OSSE P $^d$& 1.58$\pm0.12$  & 500$^{+\infty} _{-315}$ &
8.2$^{+0.3} _{-0.5}$ & - & - & -  & 459.6/510 (0.90) \\

13 & {\it ASCA}$+$OSSE R $^c$&  1.74$^{+0.33} _{-0.30}$  & $335 
^{+\infty} _{-255}$  & 8.4$\pm$0.7 & - & - & $ 1.5 ^{+1.9} _{-1.4}$
 $^e$ &350.0/410 (0.85) \\

14 & {\it ASCA}$+$OSSE T $^c$&  1.70$^{+0.31} _{-0.28}$  & 165$^{+820}
 _{-90}$ & 7.5$^{+0.6} _{-0.7}$ & 75$^{+95} _{-6}$ $^e$ &0.43$
 \pm 0.35$ $^e$  &  - & 349.7/409 (0.86) \\

\\
\hline
\end{tabular}
\end{flushleft}

\end{table*}

Using the OSSE data from viewing period 411.5,
the normalizations for the GIS respectively the OSSE data differ 
by a factor of 2, with the GIS data being stronger and there is no 
evidence for a cut-off. Renormalizing the GIS 
data to the normalization of the OSSE data, gives a flux $F_{2-10} = 2.2
\times 10^{-11}$ erg cm$^{-2}$ s$^{-1}$  and a subsequent fit, keeping 
the GIS band fixed,
still does not need a cut-off. This could also have been suspected in the 
OSSE data themselves, which shows a flat spectrum during that 
viewing period. Figure 3 shows the combined GIS and  OSSE data 
from VP 411.5, and the best-fit, unfolded model, plotted as  
$E  F _E$ versus $E$. The difference in normalization is clearly seen.

\begin{figure}
\psfig{figure=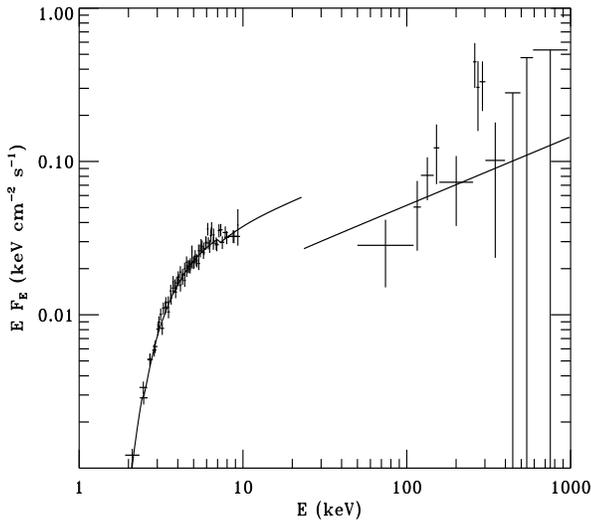,height=6.8truecm,width=8.7truecm,angle=0}
\caption[ ]{ An $E  F _E$ spectral plot of the combined {\it ASCA} 
GIS/{\it CGRO} OSSE data, from the viewing period 411.5, of NGC 7172.
The solid curve shows the best fit
`model P' (power law with an exponential cut-off and 
a neutral absorber). For best fit values, see text, or 
row 9 in table 3}
\end{figure}

The difference in the normalization between the GIS and the OSSE is also 
obvious, when the VP 412.0 data are used, with the GIS data being 
$\sim 30 \%$ lower. The data now clearly show a sharp cut-off at 
$E_{\rm fold} = 45 ^{+510} _{-25}$ keV, which already is indicated in the OSSE 
data alone, exhibiting a steep power law. The photon index of the total fit 
is also affected, as it gets somewhat flatter; $\Gamma = 1.44 ^{+0.19}
 _{-0.26}$.
Figure 4 shows the combined GIS 
and  OSSE data from VP 412.0, and the best-fit, unfolded model,
plotted as $E F _E$ versus $E$.

\begin{figure}
\psfig{figure=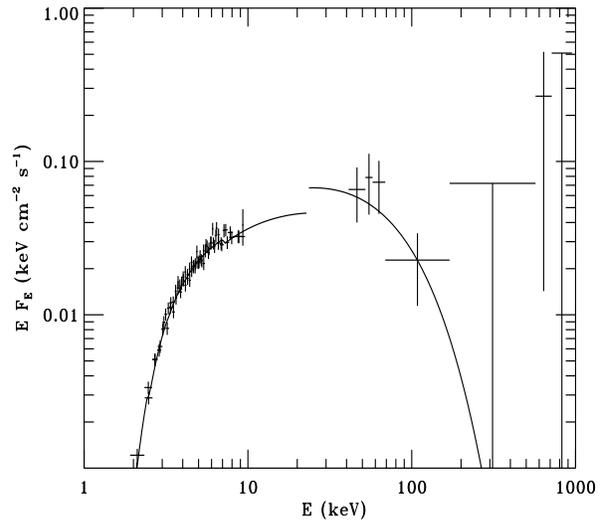,height=6.8truecm,width=8.7truecm,angle=0}
\caption[ ]{Same as Fig. 3., but for viewing period 412.0.
For best fit values, see text, or 
row 10 in table 3}
\end{figure}

In the case of using the VP 413.0 data, there is no 
difference in normalization ($< 5 \%$) between the GIS and the OSSE 
ranges and there is also no statistically significant change of the fit,
when the normalizations are locked to each other ($\Delta \chi ^2 < 0.5$).
The data also show a sharp cut-off at 
$E_{\rm fold} = 140 ^{+310} _{-70}$ keV. 

The combinations of data studied above, unequivocally indicate
that the e-folding energy has changed during the 
observation period. The photon index does not change 
significantly, as it probably,  more or less, is determined by the 
{\it ASCA} data. The OSSE data does, however, have an impact on
the fit.

Of the data studied, we find the GIS data combined with the OSSE data
from VP 413 as the most interesting combination. Firstly, the 
normalization are consistent 
with each other, secondly the VP 413 data set has the longest effective 
observation time, leading to the best quality of the data sets and 
thirdly, it can be mentioned that VP 413 is the data set 
that is closest to the {\it ASCA} observations. The resulting fit gives 
$\Gamma =  1.54 \pm 0.13$, 
and $\NH = (8.1 \pm 0.6) \times 10^{22}$ cm$^{-2}$, with  
$\chi ^2 /$d.o.f. $ = 
 351.1 / 411 $ ($\chi ^2 _{\nu} = 0.85$). The 
fitted spectrum is as flat as the one obtained with the GIS data
alone. Figure 5 shows the combined GIS 
and  OSSE data from VP 413.0, and the best-fit, unfolded model,
plotted as $E F _E$ versus $E$.

\begin{figure}
\psfig{figure=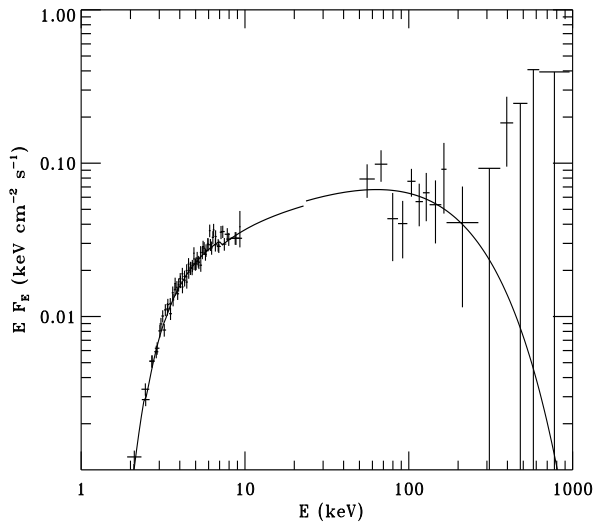,height=6.8truecm,width=8.7truecm,angle=0}
\caption[ ]{Same as Fig. 3., but for viewing period 413.0.
For best fit values, see text, or 
row 11 in table 3}
\end{figure}

We continue studying the GIS/OSSE(VP 413) data by applying more 
advanced models. We begin by trying to add a Compton
reflection component. By adding the reflection 
parameter $R$, the $\chi ^2$ of the
resulting fit is reduced by only $1.1$ units. 
Furthermore, the data can not properly constrain the 
$R$ parameter. At a 68 $\%$ confidence level $R$ lies
 between $5 \times 10^{-2}$ and 3.4,
so even at this level the resulting fit is consistent with 
no reflection.

Continuing with model T one finds a fit with a 
power law, $\Gamma =  1.70 ^{+0.31} _{-0.28}$ with $ E _{\rm fold} 
= 165 ^{+820} _{-90}$ keV and
$\NH = (7.5 ^{+0.6} _{-0.7}) \times 10^{22}$ cm$^{-2}$. 
Here too, the data have 
difficulty in constraining the model. 
At a 68 $\%$ confidence 
level the best fit has  $\NHp = (73^{+95} _{-6}) \times 10^{22}$ 
cm$^{-2}$ and $C _ F = 0.43 \pm 0.35 $. 
The fit is not better than the fit of model P, $\chi ^2 /$d.o.f. $ = 
350 / 409 $ ($\chi ^2 _{\nu} = 0.86$).
Neither the reflection model nor the transmission model is therefore 
preferred by the data over the power law model, and thus 
the OSSE data do not alter, or at least do not 
contradict, the conclusion  from using
the {\it ASCA} data alone, i.e., that the X-ray spectrum of NGC 7172  
has become flatter between 1989 and 1995.
The conclusion of the tests to prove the flatness of the spectrum 
found by the  GIS data alone, can, however,
only be tentative,
due to the spectral and 
normalization variations in the OSSE band in the short time scales, observed.

With the available data, and bearing in mind the problems connected
with non-simultaneous studies, we present the results from model P, 
using GIS data from the {\it ASCA} observation and the VP 413.0
data from the {\it CGRO} OSSE observation, GIS/OSSE(VP~413), as the
finally most interesting ones (Fig. 5.). Figure 6 shows a 
contour plot of the column density 
versus the photon index of the power law.  The contours are
90 $\%$ and 68 $\%$ confidence levels. 
The figure manifests the flatness of the spectrum.
Another contour plot is shown in Fig. 7. Here the   
photon index is plotted against the e-folding energy. The 
contours are again 90 $\%$ and 68 $\%$ confidence levels. 
The figure shows that the e-folding energy is not as tightly 
constrained to low energies, as the single parameter interval 
suggests, but it is obviously bounded upwards.

\begin{figure}
\psfig{figure=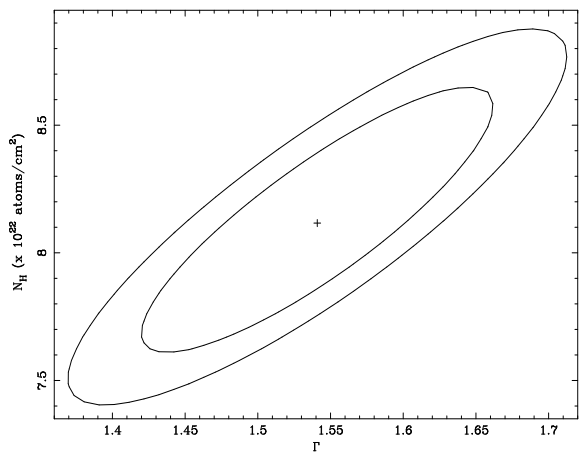,height=6.8truecm,width=8.7truecm,angle=270}
\caption[ ]{ Confidence contours for the column density, $N_{\rm H}$,
 versus the photon index, $\Gamma$, for the ASCA+OSSE(VP413) P fit.
 The contours have delta-fit
statistic of 2.30 and 4.61, i.e. 68 $\%$ and 90 $\%$ confidence
levels for two interesting parameters. The parameter values corresponding
to the best fit is marked by a plus sign ($+$)}
\end{figure}

\begin{figure}
\psfig{figure=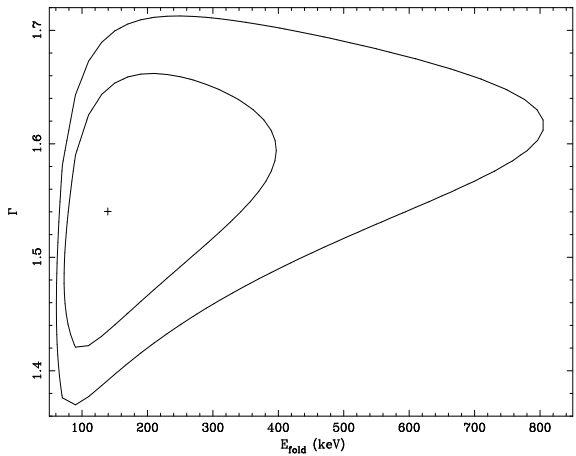,height=6.8truecm,width=8.7truecm,angle=270}
\caption[ ]{Confidence contours for the photon index, $\Gamma$, versus 
the e-folding energy, $E_{\rm fold}$, for the ASCA+OSSE(VP413) P fit. 
The contours have delta-fit statistic of 2.30 and 
4.61, i.e. 68 $\%$ and 90 $\%$ confidence levels for two interesting
 parameters. The parameter values corresponding to the best fit 
is marked by a plus sign ($+$)}
\end{figure}

 The {\it ASCA}
data below $\sim 2$ keV are not used in the modelling.
It is, however, evident that a soft ($ < 2$ keV) component
exists, which is often seen in Seyfert 2 galaxies
(Morse et~al. {\cite{Morse}}). 
In this case, the power
of this soft component is about a factor of $10^3$ lower than 
the peak power of the source.
The $0.1 - 2.4$ keV luminosity is found to be
approximately 10 times larger than the 
{\it ROSAT}/PSPC measurement on 18 November 1992  
of $L =  2.3 \times 10^{40}$ erg s$^{-1}$ (Polletta et~al. 
\cite{Polletta}). It probably originates
from a parsec sized region far away from the nuclear source.
It may be due to scattered nuclear light or thermal bremsstrahlung.

For comparison, we also study the combined {\it ASCA}/OSSE(VP 413)
spectrum using the SIS0 data instead, which is the least
conspicuous of the SIS observations 
as compared to the GIS. The resulting fit confirms the low 
e-folding energy, $E_{\rm fold} = 115 ^{+190} _{-55}$ and the 
flat power law  $\Gamma = 1.39 ^{+0.18}  _{-0.17}$.

\subsection{The combined {\it Ginga}/OSSE spectrum}

We now turn over to the {\it Ginga} data and combine it
with the OSSE data, noting 
that several years separate these two observations, which demands
even more caution to be taken, in interpreting the non-simultaneous 
fits. 
In the first fit, using the OSSE data from viewing period 411.5
the normalization of the OSSE data are only 10 $\%$ higher than 
the {\it Ginga} data. There is no cut-off at high energies.
The best fit has 
$\chi ^2 /$d.o.f. $ = 96.6 / 74 $ ($\chi ^2 _{\nu} = 1.31$)
with $\Gamma = 1.84\pm 0.04 $.

The next two fits include the OSSE data from viewing period 
412.0 and 413.0. The photon index of the power law is  
dictated by the high quality {\it Ginga} data, $\Gamma = 1.8$. 
The free normalization shows that the {\it Ginga} data  
is 50 $\%$ higher and 60 $\%$ lower, respectively. For these
two periods, the e-folding energies are low, even with
the steeper spectrum given by the {\it Ginga} data. 
They are not constrained upwards, with 90 $\%$ lower limits 
of 95 and 120 keV, respectively.

\subsection{The combined GIS/{\it Ginga}/OSSE
spectrum}
 
Finally, we analyse all three data sets together. It is evident from
an inspection of the data that there is a  difference in the
normalization between
the 1995 {\it ASCA} and the 1989 {\it Ginga} data, and we allow the 
{\it Ginga} data to have variable
normalization relative to the {\it ASCA} and OSSE data sets. 
From inspection of the residuals of the different instruments,
it is clear that the high quality {\it Ginga} data that defines
the fit, and leaving the ASCA data poorly fit. 
This supports earlier findings, that NGC
7172 had a different spectral shape and absorber in 1989 than in 1995
and we do not make any further analysis using data from all three satellites
at the same time. 

\section{Discussion}

Using the {\it ASCA} GIS observations of NGC 7172, the
best fit for a power law and a single absorber (model P) 
gives a flat spectrum with $\Gamma =
1.6$. Adding the {\it CGRO} OSSE data, a broader band 
spectrum is achieved, 
which could constrain the spectral parameters better. 
The GIS/OSSE observations of NGC 7172 were, however, not simultaneous.
The time interval between the observations was only two months and
the source did not show any evidence of variability during the 
{\it ASCA} observation. Furthermore, the flux variation over the
period 1985--1995 was quite  modest. On the other hand, the OSSE 
observations showed considerable spectral variations during the 
month it was observed. In addition, the {\it ASCA} data from 
the May 1996 observation (Matt et~al. \cite{Matt}) show
a decrease in $2 - 10$ keV flux, down to $F_{2-10} = (1.15\pm 0.05) 
\times 10^{-11}$ erg cm$^{-2}$ s$^{-1}$. This emphasises the difficulty of
combining non-simultaneous observations, which should be studied using
great caution. Very few broad-band studies have been done 
and as they are of great physical interest, it is useful to 
study even non-simultaneous data sets, taking into account the 
uncertainties that infallibly accompanies the study. 

The combined data from March and May 1995 
are consistent with the spectrum of NGC 7172 being as flat as 
observed with the {\it ASCA} satellite alone. The OSSE data show a rather
sharp cut-off at $E_{\rm fold} \sim  150$ keV, and it is relatively well
constrained.
  The fits do not improve 
when more complicated
 models are used, such as adding an Fe line, a reflection component, 
or using a dual 
absorber model.

 The flat spectrum found means
that the spectral index of NGC 7172 is variable, changing from
$\Gamma = 1.85$ in 1989 to $\Gamma = 1.5$ in 1995. It still lies 
within the range that Seyfert 1 galaxies are observed to have. 
We note that some other Seyfert galaxies (see, e.g., Yaqoob \cite{Y2}
and references therein), 
most notably 
NGC 4151 (Yaqoob \& Warwick \cite{YW}, Zdziarski et~al. \cite{Z6}), 
also have variable $\Gamma$. Furthermore, both {\it Ginga} 
(Smith \& Done \cite{SmithDone}) as well as
{\it ASCA} (e.g. Cappi et~al. \cite{Cappi}, Weaver et~al. \cite{Weaver})
observations show that some Seyfert 2 galaxies have
flat spectra similar to that sometimes shown by NGC4151.
In a large fraction of these sources with flat spectra, reflection 
is not needed in the fits, and thus the intrinsic spectrum 
could be flat.

Zdziarski et~al. (\cite{Z6}) discussed in detail the theoretical implications
of the flat intrinsic spectrum and weak reflection of NGC 4151. Similar
arguments also apply to NGC 7172. A flat intrinsic spectrum implies that the
X/$\gamma$ source is soft-photon starved. This, in turn, requires that any
cold matter reprocessing the X/$\gamma$ radiation into soft UV-photons must
subtend a small solid angle as viewed from the X/$\gamma$ source, and/or
that the X/$\gamma$ source subtends a small solid angle as viewed from the
cold matter.
Weak reflection, furthermore, requires similarly that the cold matter
subtends
a small angle as viewed from the X/$\gamma$ source, or, if the cold matter
has a disk geometry, that the disk is viewed nearly edge-on.
Several geometries are consistent with these requirements, for instance,
a hot inner disk and a cold outer disk (Shapiro et~al. \cite{SLE}),
or edge-on view of a cold disk with hot sources elevated above the disk
(e.g., Svensson \cite{S96}). 
The varying spectral slope can then be due
to a change in geometry, such as changing the interface radius between
the inner hot and the outer cold disk, or changing the elevations of the 
hot sources above the cold disk.
As we are likely to view the central source in Seyfert 2 galaxies
more or less edge-on, anisotropic
spectral effects are a less likely source of spectral variability 
(Stern et~al. \cite{Stern95}, Poutanen et~al. \cite{PSS}).

On the other hand, a reflection model or a 
transmission model with a photon index, $\Gamma = 1.7$, cannot be 
ruled out completely.
The fits are, however, not significantly better. In the case of the 
dual absorber, the geometry of the absorbing gas components with covering
fraction $\sim 0.5$ over scales of $10^{13}-10^{14}$ cm
(i.e., the scale of the central X-ray source) is not obvious.

It is of interest to compare our results with previous efforts
to fit broad band spectra of Seyfert 2 galaxies.
NGC 4945 (Done et~al. \cite{MadDone}) shows a steep intrinsic spectrum
with $\Gamma = 1.82$ and a very large absorber,
 $\NH = 4 \times 10^{24}$ cm$^{-2}$.
The e-folding energy is not well constrained at $E_{\rm fold} = 1000
 ^{+\infty} _{-730}$ keV.
Our results for NGC 7172 are more consistent with the average broad band 
spectrum of the three Seyfert 2 galaxies, 
NGC 4507, NGC 7582, and MCG-523-16 (Zdziarski et~al. \cite{Z5}) which displays
a flatter spectrum with $\Gamma = 1.67$ and an e-folding energy 
at $ E_{\rm fold} = 250 ^{+\infty} _{-180}$ keV.

\section{Conclusions}

According to the {\it ASCA} observations from 1995 and 1996, the
Seyfert 2 galaxy NGC 7172 exhibits a flat spectrum, $\Gamma \approx 1.5$,
 unlike 
the typical Seyfert 1 galaxies, thus posing a problem for the current
theory of the Seyfert unification scheme, which predicts the same average 
intrinsic photon index for Seyfert 2 galaxies  and Seyfert 1 galaxies,
i.e., a mean value of $\Gamma = 1.9 - 2.0$. 
Moreover, the 1989 observation by {\it Ginga}, with its broader spectral 
range, showed a spectral slope, $\Gamma = 1.85$, similar to that of 
Seyfert 1 galaxies. 
By combining the {\it ASCA} and the {\it CGRO} OSSE data, both from 1995, a
broad band spectral fit  is possible, which better determines
the effects of a possible complex absorber and of Compton reflection, which 
may be undetectable using the {\it ASCA} data alone. The data are, however,
non-simultaneous and the final results should therefore be treated with 
appropriate caution.
 
 We find that the simplest model, consisting of an absorbed power law with 
an exponential high energy cut-off, gives an acceptable description 
of the data.
The fit does not change significantly 
as compared to the results obtained by analysing only the {\it ASCA} data.
The spectral index of the power law has thus become flatter since 
the {\it Ginga} observations in 1989 and the change in spectral slope,
$\Delta \Gamma \approx 0.3$,
 could 
be an intrinsic change. The value obtained for the photon index still 
lies at the lower end of the range observed for Seyfert 1 galaxies.
The spectrum is observed to cut-off at about 150 keV, using the
{\it CGRO} OSSE viewing period with the highest quality data. We
also find that the {\it CGRO} OSSE spectral shape appears to vary on
a time scale of weeks. 
 
\begin{acknowledgements}

We wish to thank Drs.~Thomas~Bridgman, Christine~Done, Matteo~Guainazzi,
the referee 
Giorgio~Matt and Niel~Johnson for enlightening discussions
 and valuable comments. 
We are also indebted to the {\it CGRO} Science Support
 Center and the  {\it ASCA} GOF at Goddard Space Flight Center 
for their assistance. 
Furthermore, we wish to thank Dr.~David~Smith for providing us with the
{\it{Ginga}} data. This research was supported by the Swedish National
Space Board and the Swedish Natural Science Research Council. 
\end{acknowledgements}

\end{document}